\documentclass{desyproc}
\usepackage{graphicx}

\newcommand{\snn}{$\sqrt{s_{_{\rm NN}}}$}

\begin{document}
\title{Latest RHIC Results on Ultra-Peripheral Collisions}

\author{{\slshape Ramiro Debbe $^1$}\\[1ex]
$^1$BNL, Upton, NY, USA}

\contribID{smith\_joe}


\acronym{EDS'09} 

\maketitle

\begin{abstract}
With RHIC running in its second phase at higher luminosities the new data sets collected so far by PHENIX and STAR are allowing improvements in the study of vector meson photo-production in Ultra peripheral Collision  events in Au+Au at the highest energy. This is a brief summary of what has been accomplished so far by both collaborations.
\end{abstract}

\section{Introduction}

Glancing heavy ion collisions exchange strong photon fluxes with intensities driven by the charge of ions and the distance between then. This 
interactions turn heavy ion colliders into laboratories to study photo-production of vector meson \cite{Afanasiev:2009hy, Abelev:2007nb} 
as well as photon-photon colliders \cite{Adams:2004rz}. These ultra-peripheral collisions (UPC) need to
be discriminated from the more abundant hadronic interaction events where the ions get close enough ($b<2R_{A}$ where b is the impact parameter defined a  the distance between nuclei centers) and produce abundant number of charged particles. Such discrimination requires special triggers which at RHIC and the LHC include rapidity gaps and low multiplicity in the rapidity region of interest.
Beam backgrounds and very peripheral hadronic interactions can reduce the efficiency of these triggers. Both PHENIX and STAR have complemented their UPC 
triggers using signal from neutrons showering in hadronic Zero Degrees Calorimeters (ZDC) placed
at zero degrees with respect to the beam direction at interaction points.
The interaction between a photon and the other ion is well described as the photon fluctuates into a vector meson which then scatters off
the other ion. The scattering can be coherent where the probe sees the target as a single object or incoherent where the scattering happens at 
the level of individual nucleons. The transverse momentum of the scattered vector mesons is a good discriminator between these two interactions as the coherent scatterings accumulate at low transverse momentum, while the incoherent ones have a broad $p_{T}$ distribution.  
The PHENIX collaboration has focused in the detection of the J/$\Psi$ meson both at mid and high rapidity making use of its electromagnetic calorimeters, ring imaging detectors and the muon arms to select events with electron or muon pairs. The STAR UPC program focused on the lower mass
$\rho^{0}$ mesons but with its bigger data sets is now able to extract a robust J/$\Psi$ signal.

\section{Recent developments}

\begin{figure}[t]
\begin{center}
  \includegraphics[width=0.46\textwidth]{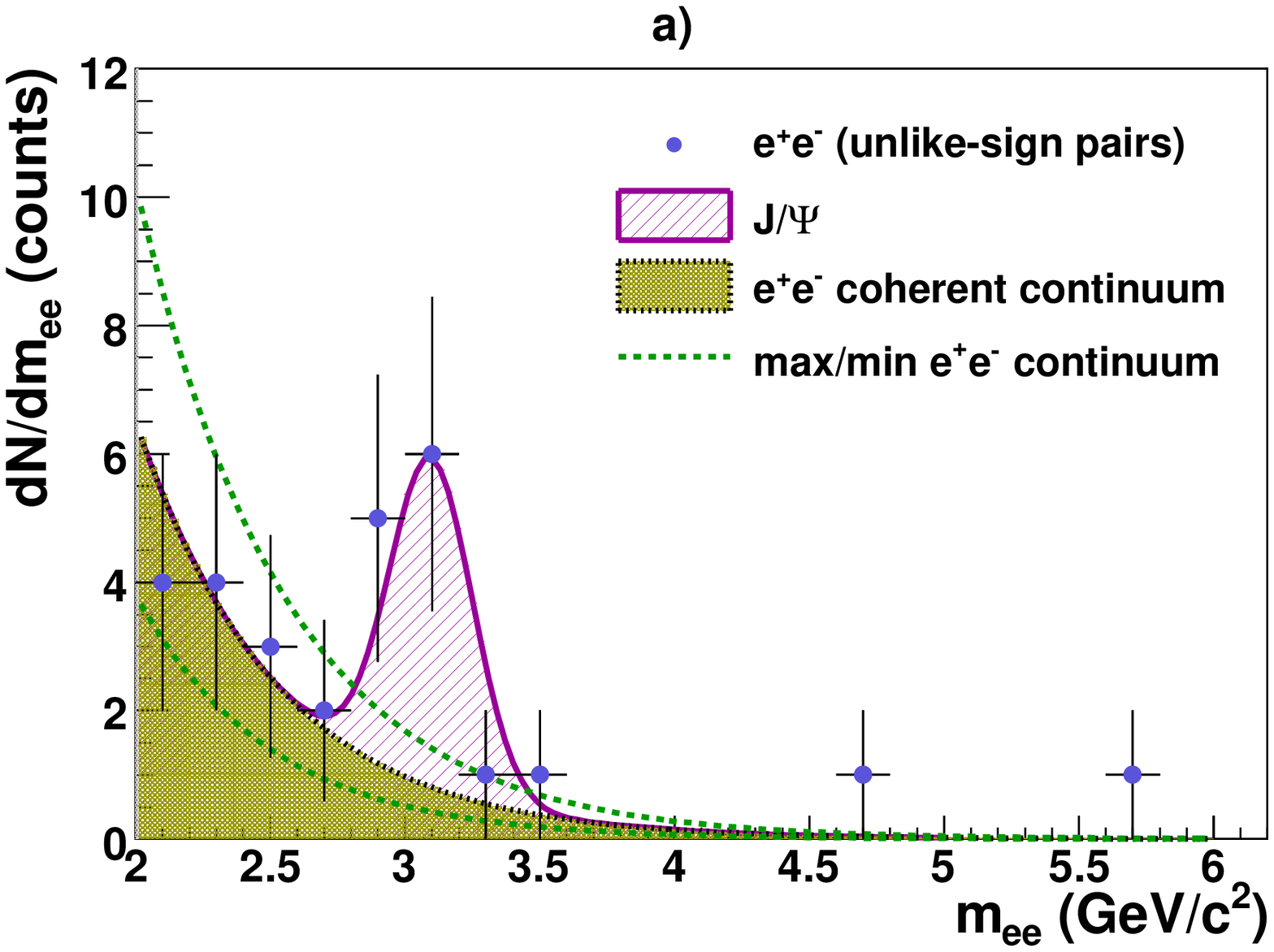}
  \includegraphics[width=0.35\textwidth]{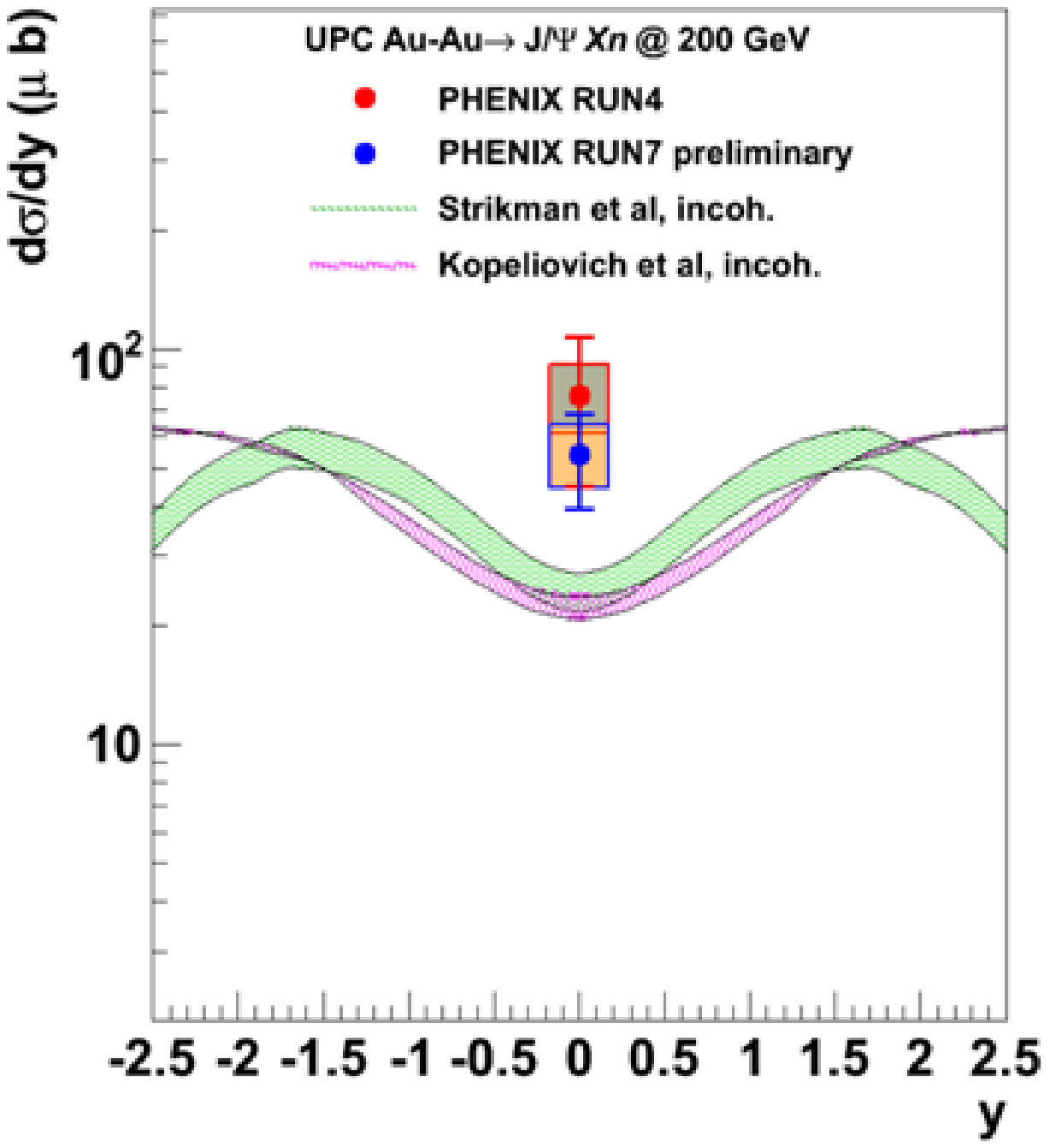}
\caption{Left, Invariant mass distribution of electron pairs detected in the PHENIX central arms. The same charge background has been
subtracted. A clear peak filled with decaying J/$\Psi$ mesons is seen as well as the $e^{+}e^{-}$ continuum. Right, cross sections for coherent 
photo-production of J/$\Psi$ at mid-rapidity. Right, the published cross section at mid-rapidity (red on line) together with the result from the
2007 data analysis.  }\label{PhenixJPsi}
\end{center}
\end{figure}

Phenix has analyzed  data collected in the years 2007 and 2010 from Au+Au interactions at \snn =200 Gev.  The 2007 data focuses 
on J/$\Psi$ mesons produced at mid-rapidity that decay into $e^{+}e^{-}$ pairs detected in the central arms. Those events include
as well the emission of at least one beam neutron which is detected in any of the ZDC detectors.
Figure \ref{PhenixJPsi} shows the background subtracted invariant mass distribution of electron pairs where the J/$\Psi$ meson peak is 
clearly visible over the di-electron continuum. The events shown in that figure have the transverse momentum of the electron pair forming 
a well defined peak below $p_{T}\sim150$ MeV/c, a clear  indication that they scattered off the entire Au nucleus. The  cross section 
extracted from 
the measurement performed during the 2004 Au+Au run is equal to $76 \pm 31 (stat) \pm 15(syst) \mu b$. The right panel of  
Fig. \ref{PhenixJPsi} shows the 2004 and 2007 cross section for J/$\Psi$ coherent photo-production at mid-rapidity. The new (preliminary) value
is $61.8 \pm 17 (stat) \pm 8.8 (syst) \mu b$ which puts the combined cross section closer to some of the model calculations. 

The STAR collaboration has a wider rapidity coverage determined by the TPC acceptance $|\eta|<1.5$ and its UPC program started with an emphasis
on the study of the photo-production of the $\rho^{0}$ meson and low mass electron pair production in $\gamma + \gamma$ interactions 
\cite{Abelev:2007nb, Adams:2004rz}. 

\begin{figure}[!hb]
\begin{center}
\includegraphics[width=0.45\textwidth]{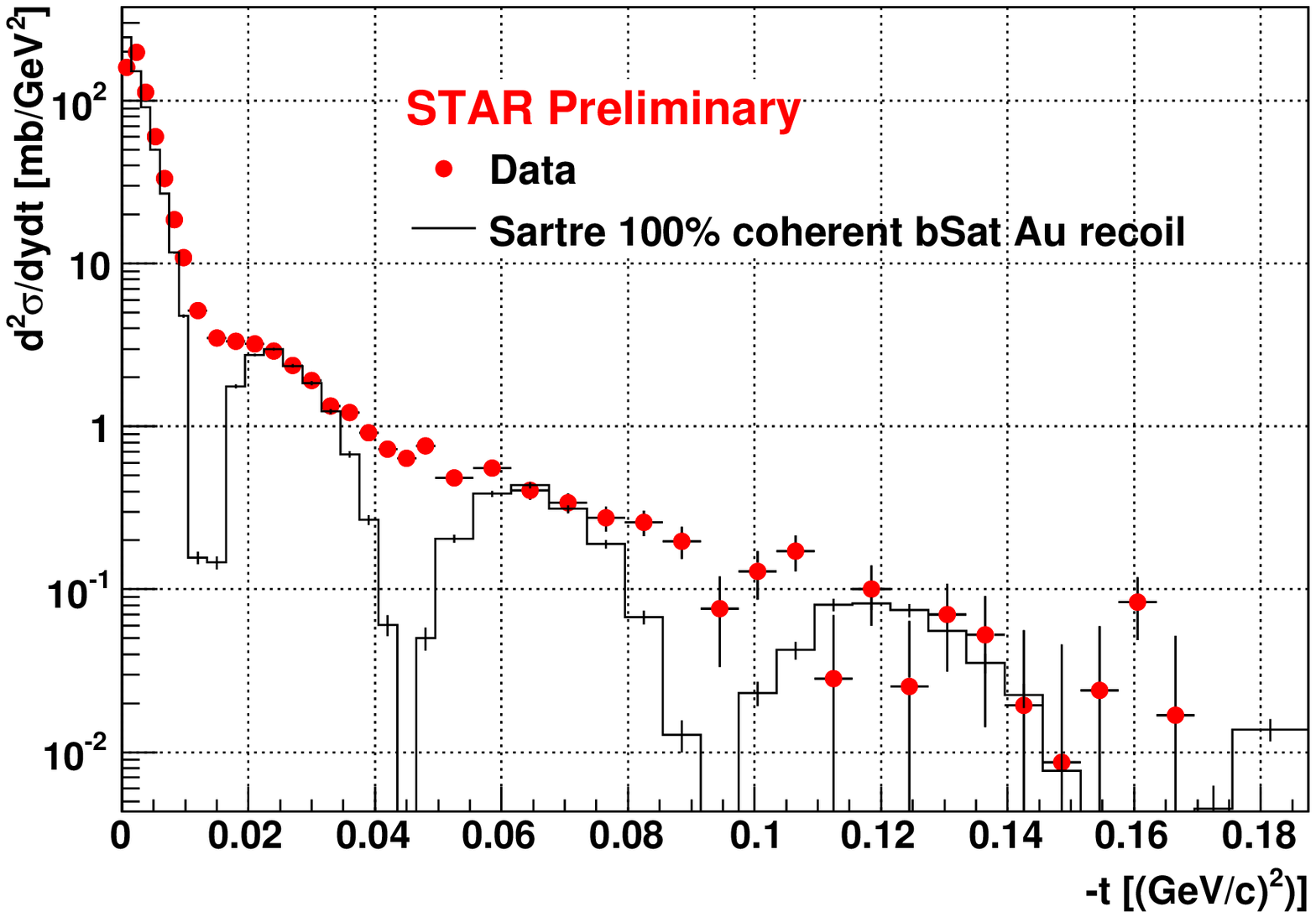}
\includegraphics[width=0.45\textwidth]{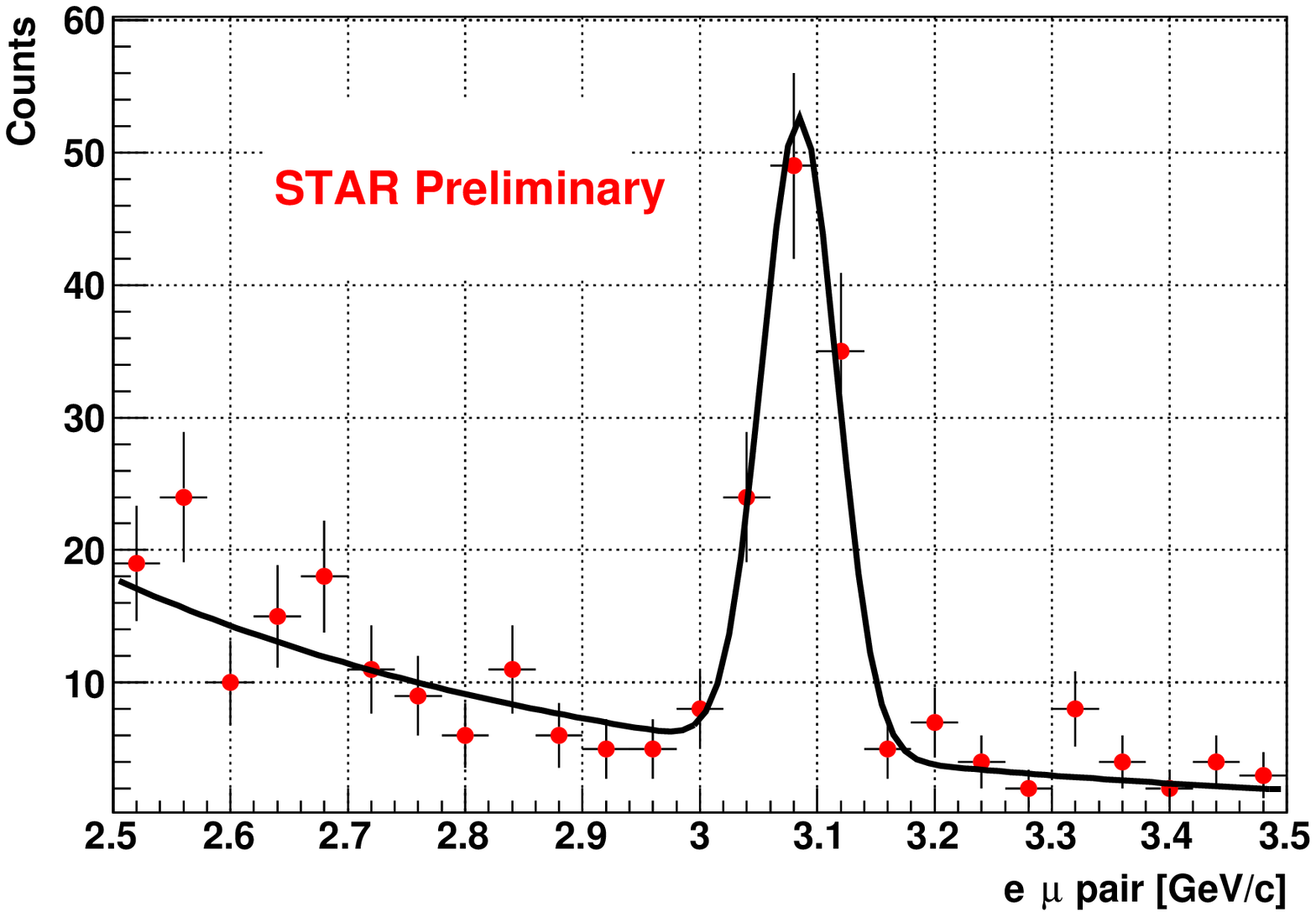}
\caption{Left, normalized distribution of the t value of the $\rho^{0}$ measured by STAR in $|y|<1$. The pion pairs originate from two track only vertices. The histogram shows the distribution of the t values for the recoiling Au nucleus generated with the Sartre event generator \cite{Toll:2012mb}. Right, invariant mass distribution of electron or muon pairs of opposite charge. Backgrounds  have been subtracted. 
}\label{rhoDiff}
\end{center}
\end{figure}

During the 2010 run which collided Au ions at \snn = 200 GeV a data set of 30M events was collected with
two triggers. Both have rapidity gaps of similar widths on both sides but they differ on the conditions at mid-rapidity. The so called UPC\_Topo
trigger uses the azimuthal information from elements that fired in the Time of Flight (TOF) detector to select back-to-back tracks. The second 
trigger called UPC\_Main demands a minimum and a maximum number of hits in the TOF detector ($2 \leq $ tof hits $\leq 6$) and requires signal 
on the ZDC detectors equivalent to $1 \leq$ beam neutrons $\leq 6$. The analyses presented in this report have been done with events that 
fire this second trigger. Events with only two tracks originating from the vertex are selected. Both tracks in those pairs have to satisfy 
quality cuts and be identified as pions: the amount of ionization energy loss in the TPC gas along the track must be  within 3 standard 
deviations away from the calculated value. The invariant mass of pair candidates with tracks having opposite charge are generated and pairs 
with masses between 0.500 and 1 GeV are selected as $\rho^{0}$ mesons. It is estimated that most of the background in this measurement comes 
from very peripheral hadronic interaction as the Au ions start to overlap. For these events with at most two tracks, the best estimator of 
the background and its relevant distributions is obtained by generating
them with pairs of equal charge. Having those equal sign distribution the background is statistically subtracted.
The coherent scattering off a nucleus is mainly concentrated at small angles (measured with respect to the photon direction which is mostly parallel to the ion beams. The incoherent component is much harder and has a power law shape. The tail of the  background subtracted
 t ($t \sim -p_{T}^{2}$) distribution is fitted with a power law function and then subtracted from the data to visualize the full extent of the coherent component. 
The left panel of Fig. \ref{rhoDiff} the result of both subtractions. For the first time in a UPC environment a diffraction pattern is
seen that is compatible with photons scattering off an object with a radius comparable to that of the Au ion~\cite{Debbe:2012aa}.

Within the same 2010 UPC data set di-electron and di-muon events were extract and J/$\Psi$ meson candidates were identified using the
value of their invariant mass. The right panel of Fig. \ref{rhoDiff} shows the distribution of the dilepton pairs after background subtraction. As was
the case for the $\rho^{0}$ meson analysis the best estimator of the background are the corresponding distributions of equal charge sign pairs.
The events that fill the histogram shown in this figure have a strong condition that demands that only two tracks originate from the vertex,
making these events candidate for exclusively photo-produced J/$\Psi$ mesons that scattered coherently off the entire Au nucleus.




\begin{footnotesize}

\end{footnotesize}
\end{document}